# 21 Million Opportunities: A 19 Facility Investigation of Factors Affecting Hand Hygiene Compliance via Linear Predictive Models

Michael T. Lash MS · Jason Slater BS · Philip M. Polgreen MD MPH · Alberto M. Segre PhD

**Acknowledgements:** The authors would like to thank GOJO Industries, Inc. for access to the hand-hygiene data.

Michael T. Lash
Department of Computer Science, University of Iowa, 319-335-0808, E-mail: michael-lash@uiowa.edu

Jason Slater
GOJO Industries, Inc.

Philip M. Polgreen
Department of Epidemiology, University of Iowa

Alberto M. Segre
Department of Computer Science, University of Iowa

**Abstract** This large-scale study, consisting of 21.3 million hand hygiene opportunities from 19 distinct facilities in 10 different states, uses linear predictive models to expose factors that may affect hand hygiene compliance. We examine the use of features such as temperature, relative humidity, influenza severity, day/night shift, federal holidays and the presence of new medical residents in predicting daily hand hygiene compliance; the investigation is undertaken using both a "global" model to glean general trends, and facility-specific models to elicit facility-specific insights. The results suggest that colder temperatures and federal holidays have an adverse effect on hand hygiene compliance rates, and that individual cultures and attitudes regarding hand hygiene exist among facilities.

## 1 Introduction

Healthcare associated infections represent a major cause of morbidity and mortality in the United States and other countries [1]. Although many can be treated, these infections add greatly to healthcare costs [2]. Furthermore, the emergence of multidrug resistant bacteria have greatly complicated treatment of healthcare associated infections [3], making the prevention of these infections even more important. One of the most effective interventions for preventing healthcare associated infections is hand hygiene [4]. Yet, despite international programs aimed at increasing hand hygiene [4, 5, 6], rates remain low, less than 50% in most cases [4, 6, 7].

Because of the importance of hand hygiene in preventing healthcare associated infections, infection control programs are encouraged to monitor rates to encourage process improvement [6, 8, 9]. In most cases, hand hygiene monitoring is done exclusively by human observers, which are still considered the gold standard for monitoring [7]. Yet, human observations are subject to a number of limitations. For example, human observers incur high costs and there are difficulties in standardizing the elicited observations. Also, the timing and location of observers can greatly affect the diversity and the quantity of observations [10, 11]. Furthermore, the distance of observers to healthcare workers under observation and the relative busyness of clinical units can adversely affect the accuracy of human observers [11]. The presence of human observers may artificially increase hand hygiene rates temporarily just as the presence of other healthcare workers can induce peer effects to increase rates [12, 13]. Finally, the number of human observations possible is quite small in comparison to the number of opportunities [7, 12].

As a consequence, several automated approaches to monitoring have been proposed [8, 14, 15, 16]. Many of these measure hand hygiene upon entering and leaving a patient's room. The subsequent activation of a nearby hand hygiene dispenser is recorded as a hand hygiene opportunity fulfilled whereas, if no such activation is observed, the opportunity is not satisfied. Such approaches, while not capturing all five moments of hand hygiene, do provide an easy and convenient measure of hand hygiene compliance. With automated approaches becoming more common, a more ongoing and comprehensive picture of hand hygiene adherence should emerge, providing new insights into why healthcare workers abstain from practicing hand hygiene.

In this work (an extension of [17]), we provide an in-depth exploration of factors affecting hand hygiene compliance across multiple hospital facilities using linear predictive models.

## 2 Data and Methods

### 2.1 Hand Hygiene Event Data

Our hand hygiene event data is a proprietary dataset provided by Gojo Industries. The data were obtained from a number of installations consisting of *door counter sensors*[1], which increment a counter anytime an individual goes in or out of a room, and *hand hygiene sensors*, which increment a counter when soap or alcohol rub are dispensed. Additional supporting technology was also installed to collect and record timestamped sensor-reported counts. We provide a simple illustration of how these technologies are used in Figure 1 and a picture of an instrumented room entrance in Figure 2. In this paper, we will use the term *dispenser event* to designate triggering and use of an instrumented hand hygiene dispenser and *door event* to designate the triggering of a counter sensor located on one of the instrumented doors.

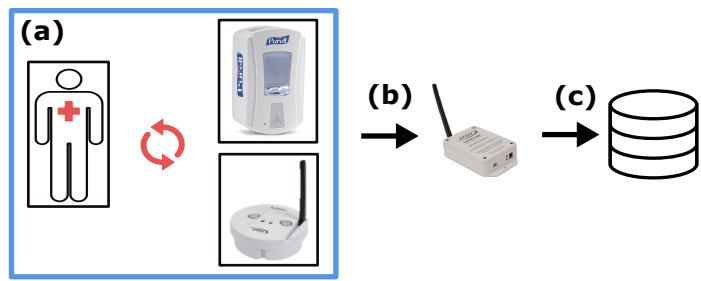

Fig. 1: A simple illustration of the sensors and corresponding infrastructure. In (a), healthcare workers enter and exit patient rooms that are fitted with sensors, interacting with instrumented dispensers as they do; note that the sensor on the hand hygiene dispenser is internal, and not visible. In (b), these door and dispenser counts are intermittently sent to a wireless transmitter. In (c), these counts are relayed via transmitter and stored in a database, along with other information, such as the room the counts came from and the time and date in which they were sent.

---

[1] Practically speaking, these sensors can be fit to any sort of patient entrance/exit area, as depicted in Figure 2.

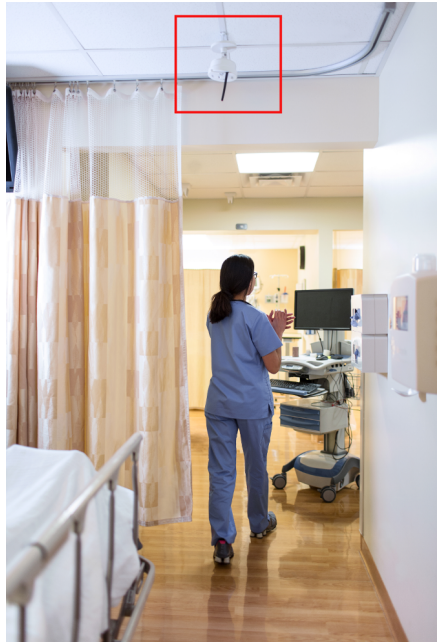

Fig. 2: A nurse applying hand hygiene rub upon leaving an instrumented patient area. Note the door sensor highlighted by the red box.

A total of 19 facilities in 10 states were outfitted with sensors; because of privacy concerns, we report only the state and CDC Division for each. The facilities comprise a wide range of geographies, spanning both coasts, the midwest, and the south. A total of 1851 door sensors and 639 dispenser sensors reported a total of 24,525,806 door events and 6,140,067 dispenser events across these 19 facilities between October 21, 2013 and July 7, 2014. Each facility contributed an average of 172.3 *reporting days*, making this study the largest investigation of hand hygiene compliance to date (i.e., larger than the 13.1 million opportunities reported in [18]). Assuming each door event corresponds to a hand hygiene opportunity, we estimate an average facility compliance rate of 25.03%, in line with if not just below the reported low-end rate found in [19].

The original data, consisting of timestamped counts reported from individual sensors over short intervals, were re-factored to support our analysis. First, data from each sensor were binned by timestamp, $t$, into 12 hour intervals, corresponding to traditional day and night shifts, as indicated by an additional variable, *night*, defined as follows:

$$nightShift = \begin{cases} 1 & t \ \epsilon \ [\text{7pm, 7am}) \\ 0 & t \ \epsilon \ [\text{7am, 7pm}) \end{cases}$$

Second, door and dispenser counts were aggregated based on day and night shift so as to produce a series of shift-level records. For each such record we compute *hand hygiene compliance*, or just *compliance*, by dividing the number of reported dispensed events by the number of door events:

$$compliance = \frac{\#\ dispenser}{\#\ door}$$

Such a definition of compliance assumes that each door event corresponds to a single *hand-hygiene opportunity* and each dispenser event corresponds to a single *hand-hygiene event* whereas, in reality, a health care worker might well be expected to perform hand hygiene more than once per entry, resulting in rates that exceed one, if only slightly. This estimator also ignores the placement of doors with respect to dispensers: multiple dispensers may well be associated with a single doorway, and some dispensers may be in rooms having multiple doors. Thus, simply adding new dispensers will raise apparent compliance rates computed in this fashion, while adding new door sensors will appear to reduce compliance. Even so, when applied consistently and if system layouts are fixed, this estimator is a reasonable approximation of true hand hygiene compliance, and supports sound comparisons within a facility (but not across facilities).

Because malfunctioning sensors or dead batteries can produce outliers (i.e., very low or very high values), shifts with fewer than 10 door or dispenser events reported per day (possibly indicating an installation undergoing maintenance), zero compliance, or compliance values greater than 1 were removed prior to analysis (at the cost of possibly excluding some legal records). The remaining data consist of 5308 shifts from the original 5647 records, having 21,273,980 hand hygiene opportunities and 5,296,749 hand hygiene events (see Table 1).

| Facility | State | CDC Div | Tot Disp | Tot Door | Days Rep |
|----------|-------|---------|----------|----------|----------|
| 91 | OH | ENC | 234292 | 518772 | 252 |
| 101 | OH | ENC | 350901 | 2021665 | 260 |
| 105 | TX | WSC | 238899 | 1940024 | 260 |
| 119 | MN | WNC | 123877 | 242939 | 156 |
| 123 | TX | WSC | 325618 | 1112198 | 243 |
| 127 | NM | Mnt | 1306855 | 4546171 | 260 |
| 135 | OH | ENC | 125731 | 264331 | 258 |
| 144 | CA | Pac | 398961 | 1744642 | 260 |
| 145 | CA | Pac | 567096 | 2073566 | 260 |
| 147 | CA | Pac | 500979 | 2462900 | 260 |
| 149 | CA | Pac | 590708 | 2306392 | 260 |
| 153 | CT | New E | 169564 | 603482 | 208 |
| 155 | NY | M-At | 171275 | 619507 | 117 |
| 156 | NC | S-At | 4381 | 38200 | 15 |
| 157 | OH | ENC | 39455 | 313396 | 101 |
| 163 | OH | ENC | 344 | 10233 | 5 |
| 168 | PA | M-At | 30421 | 86909 | 20 |
| 170 | IL | ENC | 112604 | 353631 | 47 |
| 173 | OH | ENC | 4788 | 15122 | 32 |
| Total | 10 | 8 | 5296749 | 21273980 | 3274 |

Table 1: Descriptive statistics for all reporting facilities in terms of state, CDC division, hand hygiene events, people events, and reporting days.

2.2 Feature Definitions

In this subsection we define the features (factors) that will be examined, and how each is derived.

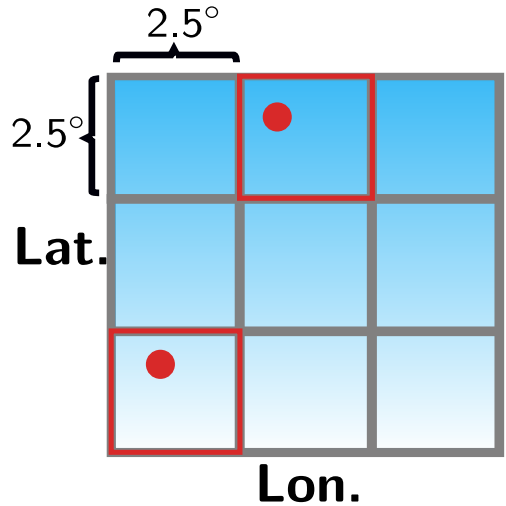

Fig. 3: Assigning (red box) NOAA weather data, reported in terms of a geographic grid, to health care facilities (red dots), where the blue color gradient might represent temperature.

2.2.1 Local Weather Data

Because health care workers frequently cite skin dryness and irritation as a factor in decreased compliance (particularly in cold weather months where environmental humidity is reduced), we associate daily air temperature (denoted $temp$) and relative humidity (denoted $humid$) to each timestamped record based on each facility's reported zip code. Spatially assimilated weather values ($\sigma = 0.995$) for the entire globe were obtained from the National Oceanic and Atmospheric Administration (NOAA) [20]. Given in terms of grid elements (a tessilation of bounding boxes covering 2.5° latitude by 2.5° longitude), the world is thus defined as a 144 by 73 grid having 10512 distinct grid elements. Weather data are available at a fine level of temporal granularity (on the order of 4 times daily for each grid unit) for the entire period of interest. The geographical assignment of weather data was obtained by first mapping each facility's numerical zipcode to the zipcode's centroid (2010 US Census data), and then subsequently mapping zipcode centroid (lat,lon) to the corresponding NOAA grid element. An example of this assignment can be observed in Figure 3. We associate weather information from the observation temporally closest to the start of each shift.

*2.2.2 Influenza Severity*

We conjecture that the local *severity* of common seasonal infectious diseases such as influenza may also affect hand hygiene compliance rates. We define influenza severity (denoted $flu$) as the number of influenza-related deaths relative to all deaths over a specified time interval.

Influenza severity data were obtained from the CDC's *Morbidity and Mortality Weekly Report* (MMWR), which also reports data at weekly temporal granularity. Rather than reporting data by CDC region, however, data are provided by *reporting city* (one of 122 participating cities, mostly large metropolitan areas). We map each facility in our dataset to the closest reporting city in order to associate the appropriate severity value to each record. In other words

$$repCity = \text{argmin}\{\text{dist}(\text{facility}, \text{city}_i) : i = 1, \ldots, 122\}$$

where $\text{dist}(\text{fac}, \text{city}) \triangleq \|(\text{fac}_{lat}, \text{fac}_{lon}), (\text{city}_{lat}, \text{city}_{lon})\|_2$, the Euclidean distance between two entities given in terms of (lat, lon) coordinates. Eight of 19 facilities were located in a reporting city (i.e., dist(fac,city)= 0). The remaining 11 facilities were mapped to a reporting city that was, on average, 66.2 miles away (only 3 of 19 facilities were mapped to a reporting city further than this average, with the largest distance being 142 miles).

*2.2.3 Temporal Factors*

We also conjecture that external factors associated with specific holidays or events may affect hand hygiene compliance rates. Holidays may change staffing rates or affect healthcare worker behaviors. The number of visitors (affecting door counter rates) may also be greater than during regular weekdays. Holidays such as the 4th of July are often associated with alcohol-related accidents, and may increase health care facility workloads (similar factors may also apply on weekends).

We define a new variable *holiday* that reflects whether a given shift occurs on one of the 10 federal holidays (New Year's Eve, Martin Luther King Day, President's Day, Memorial Day, the 4th of July, Labor Day, Columbus Day, Veteran's Day, Thanksgiving or Christmas) where, if any part of the shift (day/night) falls on the holiday in question, the indicator is set to 1. More formally:

$$holiday = \begin{cases} 0 & t \notin \{holidays\} \\ 1 & t \in \{holidays\} \end{cases}$$

Similarly, in order to ascertain the impact of weekends on compliance, we define a new variable *weekday* as follows:

$$weekday = \begin{cases} 0 & t \in \{Sat, Sun\} \\ 1 & t \in \{Mon, Tues, Weds, Thurs, Fri\} \end{cases}$$

Note here that if a shift spans the weekday into a weekend (or vice versa), it is encoded as a weekend.

A related concept is the presence of new resident physicians, who traditionally start work the first of July. We define a new variable that corresponds with this time period in order to see if the data reveal the presence of a July effect (denoted *July*):

$$July = \begin{cases} 0 & t \notin July_{1-7} \\ 1 & t \in July_{1-7} \end{cases}$$

2.3 Exploring Factors Affecting Hand Hygiene

*2.3.1 M5 Ridge Regression for Feature Examination*

With covariates defined and associated with the collected sensor data, we wish to build a linear hypothesis **h** that **(a)** accurately estimates hand hygiene and **(b)** reports the direction and degree of effect of our defined features.

In accomplishing **(b)** we bear in mind two things:

**(1)** There may be multi-collinearity among features, which may adversely affect the output.
**(2)** That **(a)** and **(b)** may be at odds with one another; i.e., obtaining good predictions may entail discarding some prediction-inhibiting features for which we would like to obtain effect estimates (in practice, we find that this is not actually the case).

Therefore, we propose an *M5 Ridge Regression for Feature Examination* method designed to accomplish **(a)** and **(b)**, while bearing **(1)** and **(2)** in mind. This method is given by

$$\mathbf{h}^* = \underset{\mathbf{h} \in \mathcal{H}_l}{\operatorname{argmin}} \; \|\Lambda(\mathbf{X})\mathbf{h} - \mathbf{y}\|_2^2 + \lambda \|\mathbf{h}\|_2^2 \qquad (1)$$
$$\text{s.t.} \quad \rho(h_j) \leq .05 \; \forall \; j$$

where $\mathbf{X} \in \mathbb{R}^{n \times p}$ is a design matrix, **h** is the hypothesis, **y** is the target vector consisting of compliance rates in which a particular $y_i \in [0,1]$, $\lambda$ is a regularization term, $\|\cdot\|_2$ is the $\ell_2$-norm, and $\rho(\cdot)$ is a function that reports the p-value of a hypothesis term (this constraint is ensured via sequential backwards elimination [21]). The function $\Lambda(\mathbf{X})$ can be defined as

$$\Lambda(\mathbf{X}) \triangleq \operatorname{argmin}\{\mathbf{t} \in T_{\mathcal{H}_l}\} \qquad (2)$$

where **t** is hypothesis selected from a tree of hypotheses constructed using the *M5* method [22]. Effectively, (2) only reduces the $p$ dimension, acting as a feature selection method, and having no bearing on the $n$ dimension.

There are a few benefits of the above method worth noting. First, the hypothesis class $\mathcal{H}_l$ is linear and common to both (1) and (2). Such two-stage optimization approaches, where the first objective is optimized taking into account the hypothesis class before the hypothesis itself is optimized for predictive accuracy (or some other such measure), have been shown to work well in other contexts [23]. Secondly, such a method is specifically geared toward producing a hypothesis that makes use of features that have an immediate bearing upon the problem, while eliminating interpretability obscuring effects, such as multi-collinearity. Moreover, these desirables are obtained while attempting to produce the most accurate hypothesis: an **h** that elicits feature indicativeness, produces accurate results, and controls for confounding effects is the goal of this two-step optimization procedure.

Ultimately, we conduct our analysis by observing the sign and magnitude of the values in the hypothesis vector in order to determine the factors that influence hand hygiene compliance, and whether such factors affect compliance in a positive or negative manner. We also observe correlation and RMSE values to determine how well our predictive model works, and whether the corresponding results can be trusted. All results and are obtained via $k$-fold cross-validation ($k = 10$).

*2.3.2 Supporting Methodology*

We also use two established/standard techniques – RReliefF feature ranking and marginal effects modeling – that will serve as a point of comparison between our method, and also help inform the discussion of the results obtained[2].

**Feature ranking:** First, we propose the use of the RReliefF algorithm [26], a modification of the original Relief algorithm of Kira and Rendell [27]. RReliefF finds a feature $j$'s weight by randomly selecting a seed instance $\mathbf{x}_i$ from design matrix $\mathbf{X}$ and then using that instance's $k$ nearest neighbors to update the attribute. This description consists of three terms: the probability of observing a different rate of hand hygiene compliance than that of the current value given that of the nearest neighbors, given by

$$A = p(\text{rate} \neq \text{rate}_{x_i} | k\text{NN}(x_i)), \quad (3)$$

the probability of observing the current attribute value given the nearest neighbors, given by

$$B = p(x_{i,j} | k\text{NN}(x_i)), \quad (4)$$

and the probability of observing a different hand hygiene rate than the current value given a different feature value $v$ and the nearest neighbors, given by

$$C = p(\text{rate} \neq \text{rate}_{x_i} | k\text{NN}(x_i) \wedge j = v). \quad (5)$$

Attribute distance weighting is used in order to place greater emphasis on instances that are closer to the seed instance when updating each term; final weights are obtained by applying Bayes' rule to the three terms maintained for each attribute, which can be expressed

$$\frac{C * B}{A} - \frac{(1 - C) * B}{1 - A}. \quad (6)$$

By using this method we could then rank attributes in terms of their importance. We again report rankings using $k$-fold ($k = 10$) cross validation.

**Marginal Effects Modeling:** To provide additional insight into the features that are relevant to hand hygiene we analyzed their marginal effects [28]. Marginal effects, also referred to as *instantaneous rates of change*, are computed by first training a hypothesis $h$, then, using the testing data, the effect of each covariate can be estimated by holding all others constant and observing the predictions. Such a method can be expressed by

$$\hat{\text{rate}}_{i,j} = \mathbf{h}^\top [x_{i,j}, \bar{\mathbf{x}}_{\neq j}] \tag{7}$$

where, with a slight abuse of notation, $x_{i,j}$, the value of instance $i$'s $j$th feature, is added to the vector $\bar{\mathbf{x}}_{\neq j}$, which consists of the average of each non-$j$ feature, at the appropriate location (namely, the $j$th position). Here, the notation $\neq j$ is used to reinforce the fact that the vector of averages $\bar{\mathbf{x}}$ has it's $j$th element replaced by $x_{i,j}$. Other non-$j$ entries are given by $\bar{\mathbf{x}}_k = \mu(\mathbf{X}_k)$, for an arbitrary index position $k$.

---

[2] Note that both the LASSO [24] and Elastic Net [25] would have also made appropriate supporting methods.

## 3 Results

### 3.1 Global Model

In this section we examine the results obtained by cross-validating *global* models, where all facility records are used, and a facility-identifying feature is included.

*3.1.1 Predictive Power: $M5$ Ridge Regression*

We learned a hypothesis using all available features, including a nominalized facility identifier. Our predictive results can be observed in Table 2. We note that the RMSE is not large and the correlation is moderate, implying relatively good predictive performance.

| Measure | Value |
|---|---|
| Correlation | 0.3441 |
| RMSE | 0.1702 |

Table 2: Correlation coefficient and RMSE of cross-validated model predictions.

*3.1.2 Examining Hypothesis $\mathbf{h}^*$*

We next examine the terms of the learned hypothesis $\mathbf{h}^*$ (see Table 3). The model includes unique identifiers for all 19 facilities, 12 of which had positive corresponding values, indicating relatively higher rates of compliance. The remaining facilities' $\mathbf{h}^*$ terms had relatively small negative values, indicating lower rates of compliance. Among other features, holidays are associated with lower compliance rates and influenza severity with higher compliance. Weekdays are associated with higher compliance rates, as are higher temperatures and humidity. Interestingly, the $M5$ Ridge Regression model appears to have eliminated some holidays (Martin Luther King day, Memorial day, Labor day, Columbus day, and Thanksgiving), as well as Facility 163 (the facility with the lowest amount of hand-hygiene data). This means that these features do not contribute to hand-hygiene compliance rates in any meaningful way.

| Feature | $h_j$ |
|---|---|
| $facility^- = \{1, 105, 147, 156, 157, 170\}$ | $h_{j \in \text{Fac}-} \in [-0.103, -0.016]$ |
| $facility^+ = \{91, 119, 123, 127, 135, 144,$ $145, 149, 153, 155, 168, 173\}$ | $h_{j \in \text{Fac}+} \in [0.008, 0.261]$ |
| $temp$ | $0.022$ |
| $humid$ | $0.0079$ |
| $weekday$ | $0.0069$ |
| $night$ | $-0.0218$ |
| $holiday = \{$Indep Day, Pres. Day, Vet Day, New Year's, Christmas$\}$ | $h_{j \in \text{Hol}} \in [-0.017, -.006]$ |
| $flu$ | $0.014$ |
| $July$ | $-0.0106$ |

Table 3: Feature specific $h_j$ terms, where red highlights features with a negative association and blue highlights those with a positive association.

*3.1.3 RReliefF*

Using RReliefF we can rank features in terms of their importance in order to support and supplement the result obtained using $M5$ Ridge Regression. These results are reported in Table 4, where rankings shown are averages for 10-fold cross-validation. Note that here $facility$ was represented as a single discretely-valued feature in order to determine the importance of facility as a whole (instead of treating each facility as its own feature), as was $holiday$.

| Attribute | Avg Val | Avg Rank |
|---|---|---|
| $facility$ | $0.029(\pm.001)$ | 1 |
| $flu$ | $0.007$ | 2 |
| $temp$ | $0.005$ | $3.3(\pm 0.46)$ |
| $weekday$ | $0.002$ | 5 |
| $humid$ | $.001$ | $6.3(\pm 0.64)$ |
| $July$ | $\approx 0.0$ | $7.2(\pm 0.4)$ |
| $holiday$ | $\approx 0.0$ | $7.8(\pm 1.08)$ |
| $night$ | $\approx 0.0$ | $8.7(\pm 0.46)$ |

Table 4: RReliefF attribute weights.

*3.1.4 Marginal Effects*

The results obtained from modeling the marginal effects can be observed in Figure 4.

Figures 4a and 4b show the marginal effects of two randomly selected facilities; one identified as being associated with lower rates of compliance and one identified as having higher rates of compliance (from Table 3). Note that, because these are binary features (taking on values of either zero or one), the kernel density of the underlying data is not readily visible (unlike the other figures, which show results for non-binary features). As we can see the marginal effects support the result obtained using both $M5$ Ridge Regression and RReliefF, and also seem to suggest an even greater association between facilities and rates of compliance than was originally apparent (at least for these two facilities).

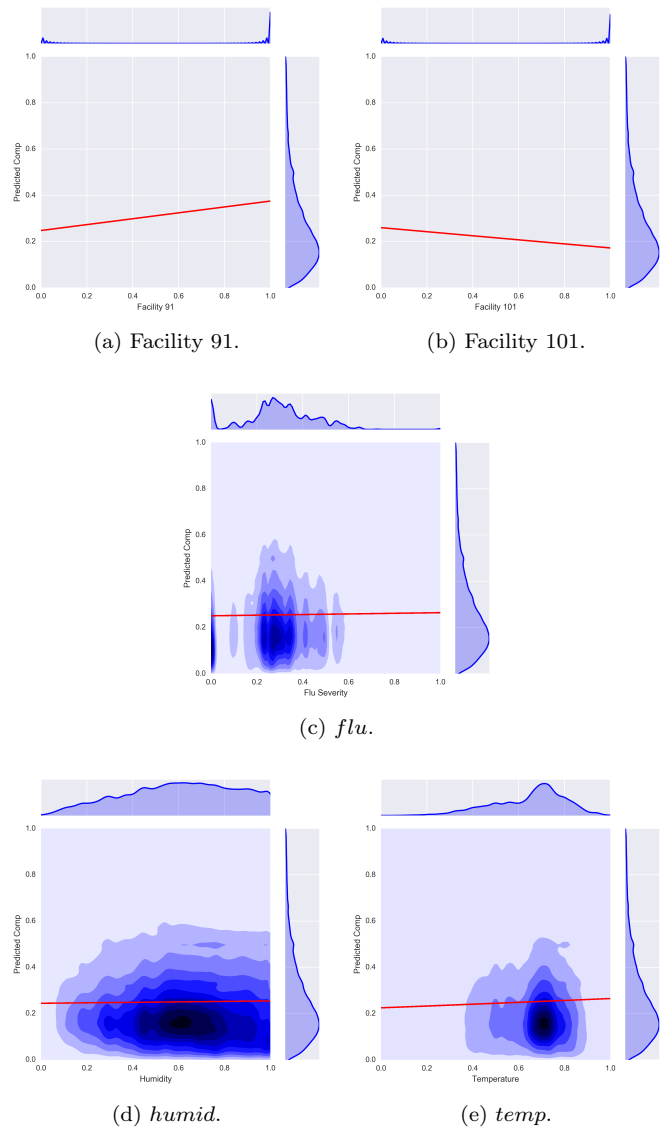

(a) Facility 91.    (b) Facility 101.

(c) *flu*.

(d) *humid*.    (e) *temp*.

Fig. 4: The marginal effects of several select covariates, where blue shows the kernel density of the original data and the red lines show the estimation. Rate (y-axis) vs. feature (x-axis). Note that in 4a and 4b no kernel density estimate is provided, as these plots are for binary features.

Figure 4c shows the marginal effects of influenza severity. The $flu$ result shows a slightly positive relationship between the severity of flu, measured in terms of mortality, and hand-hygiene compliance rates. This is further supported by the result obtained from $M5$ Ridge Regression and the RReliefF ranking.

Figures 4d and 4e show the marginal effects of humidity and temperature. The result obtained for both is consistent with that from $M5$ Ridge Regression. The lesser effect of humidity and greater effect of temperature are also reflected in the RReliefF ranking.

### 3.1.5 Weather and Temperature: Statistical Significance

To further explore the relationship between hand-hygiene and weather effects, we conducted a simple statistical analysis. For each facility, we selected the temperature and humidity values corresponding to the bottom 10% and top 10% of hand-hygiene compliance rates. We then performed a paired t-test on each set of samples; temperature and humidity values were scaled to $[0, 1]$. The results of this analysis are reported in Table 5.

| Facility | State | $temp$ $\mu_{\text{top}} - \mu_{\text{bot}}$ (p-val) | $humid$ $\mu_{\text{top}} - \mu_{\text{bot}}$ (p-val) |
|---|---|---|---|
| 91 | OH | -0.004 (0.750) | -0.007 (0.489) |
| 101 | OH | 0.001 (0.909) | 0.004 (0.457) |
| **105** | TX | 0.041 ($< 0.000$) | -0.028 (0.001) |
| 119 | MN | -0.008 (0.699) | -0.013 (0.337) |
| **123** | TX | 0.017 (0.002) | 0.029 ($< 0.000$) |
| **127** | NM | 0.032 ($< 0.000$) | -0.063 ($< 0.000$) |
| 135 | OH | -0.045 (0.010) | 0.017 (0.278) |
| **144** | CA | 0.009 ($< 0.000$) | -0.018 (0.002) |
| 145 | CA | -0.001 (0.675) | 0.004 (0.549) |
| **147** | CA | 0.011 ($< 0.000$) | -0.013 (0.017) |
| 149 | CA | -0.007 (0.025) | 0.008 (0.214) |
| **153** | CT | 0.043 ($< 0.000$) | -0.003 (0.746) |
| **155** | NY | 0.093 ($< 0.000$) | 0.012 (0.341) |
| **156** | NC | 0.040 (0.007) | -0.041 (0.445) |
| 157 | OH | -0.132 ($< 0.000$) | -0.020 (0.638) |
| **163** | OH | 0.180 (0.010) | 0.179 (0.021) |
| **168** | PA | 0.012 (0.122) | 0.071 (0.006) |
| 170 | IL | -0.001 (0.772) | -0.007 (0.642) |
| **173** | OH | 0.037 (0.003) | -0.033 (0.440) |

Table 5: The difference in means and paired t-test p-value results, obtained by comparing temperature/humidity values among the bottom 10% and top 10% of hand-hygiene compliance rates, by facility (**boldened blue** indicates that either temperature, humidity, or both have a positive difference in means and a p-value $\leq .05$).

Table 5 shows that most facilities have statistically significant differences between the two samples and that $\mu_{\text{top 10}} > \mu_{\text{bottom 10}}$. Such results indicates that higher temperatures and levels of humidity (particularly temperature) are statistically associated with higher rates of hand hygiene. However, we find that some facilities co-located in the same geographic region have conflicting statistical results (e.g., Facs. 91, 173). We conjecture that such a result may attributable to differences in sensor deployment location, but we leave such an investigation as future work.

3.2 Facility-Specific Modeling

The full $M5$ Ridge Regression models' reliance on facility identities, coupled with the RReliefF feature ranking result, suggests that compliance depends, at least in part, upon facility-specific health care worker attitudes, administrative culture, or even simply the disposition of sensors and the architecture of the facility. Therefore, we propose to construct and analyze facility-specific models in the same manner as our global model.

Therefore, in this section, we present a comprehensive set of facility-specific results obtained using facility-specific models: we explore 10 of the 19 facilities disclosed in Table 1, which comprehensively represent a large geographic dispersion (which may produce geographic-specific similarities and differences in the obtained results), which will further illustrate the facility (and location)-specific factors affecting hand hygiene compliance.

3.2.1 Predictive Power: $M5$ Ridge Regression

The facility-specific $M5$ Ridge Regression modeling results are reported in Table 6.

| Fac # | Correlation | RMSE   |
|-------|-------------|--------|
| 155   | 0.5907      | 0.0658 |
| 153   | 0.2089      | 0.0991 |
| 149   | 0.1168      | .0489  |
| 123   | 0.6193      | 0.11   |
| 127   | 0.7133      | .0313  |
| 91    | 0.5384      | 0.0939 |
| 101   | 0.3751      | 0.0442 |
| 170   | 0.0645      | 0.0607 |
| 168   | 0.362       | 0.0794 |

Table 6: Facility-specific $M5$ Ridge Regression cross-validation results.

Comparing Table 2, showing the performance of the global model, with Table 6, we can see that there is uniformly lower RMSE among the facility-specific models (Table 6) as compared to that of the global model. This result is not unexpected. On the other hand, we observe a range of correlation values, some of which are better than the global model (the first eight facilities in Table 6), and some of which that are worse. We note that the last two facilities, which had worse correlation results than the global model, are also the facilities that have comparably little data.

We now turn to examining the terms of each facility-specific hypothesis vector, which can be observed in Table 7. Note that, for the sake of simplicity in analyzing these features, we have created a single, binary *holiday* feature (as opposed to having a feature for each holiday, as in our global model).

| *facility #* | *temp* | *humid* | *weekday* | *flu* | *holiday* | *night* | *July* |
|---|---|---|---|---|---|---|---|
| 147 | 0.4237 | 0.0594 | NA | $-0.937$ | NA | $-0.0176$ | NA |
| 155 | 0.2721 | NA | 0.0491 | 0.1847 | NA | -0.178 | NA |
| 153 | NA | 0.048 | 0.0168 | -0.0638 | NA | -0.0514 | -0.0779 |
| 149 | NA | NA | 0.0184 | -0.0543 | NA | 0.0093 | NA |
| 123 | 0.419 | NA | -0.0572 | -0.2392 | NA | 0.0787 | NA |
| 127 | 0.0672 | NA | 0.0315 | 0.0383 | -0.0232 | -0.0499 | NA |
| 91 | -0.0546 | 0.1329 | 0.0683 | NA | -0.1207 | -0.1012 | NA |
| 101 | -0.0437 | 0.0219 | 0.0219 | NA | -0.0234 | -0.0169 | -0.0617 |
| 170 | NA | NA | NA | -0.1518 | NA | NA | NA |
| 168 | NA | 0.1414 | NA | 0.207 | NA | -0.0742 | -0.0729 |

Table 7: Hypothesis vector terms for each facility-specific model.

In examining Table 7, we wish to first point out that, relative to the global model result reported in Table 3, that all facility-specific models had at least one term that was removed via sequential backwards elimination. Moreover, these eliminated terms differ by facility, demonstrating that local models are sensitive to different features in different ways.

In examining the hypothesis terms, some interesting findings emerge. With respect to our weather-based features – temperature and humidity – we can see that, for the most part, these factors were positively associated with higher rates of hand hygiene compliance and, for certain facilities (147, 155, 123), these features appear to be fairly important (based on the magnitude of the coefficients). Two facilities, however, have a negative association with temperature and compliance. These coefficients, however, are relatively small and are offset by positive associations among humidity: in other words, the effects of temperature on compliance rates at these facilities appear to be somewhat negligible.

In examining *weekday* and *holiday*, we can see that in all but one facility, *weekday* has a positive influence on hand hygiene rates. This suggests that employees that work during weekends at these facilities may be washing their hands less; this may be attributable to a number of factors (increased work load, etc.). The *holiday* feature, on the other hand, tends to be indicative of lower rates of compliance among the three facilities reporting a non-zero term in their hypothesis vector (i.e., facilities 91, 101, 127).

The *night* and *July* features also tend to be negatively associated with hand hygiene compliance, with $JulyEffect$ being universally associated with negative rates of compliance (among the three facilities for which this term was not eliminated). *night*, by contrast, had two facilities which were found to have a positive term for this feature. These may be hospitals where there is relatively less activity at night (less busy); however, further investigation is needed to tease out the reasons individual facilities experience these differing rates.

Finally, *flu* appears to have a mix of positive and negative associations among facilities. In those facilities that have negative associations, a campaign focusing on flu awareness may be beneficial; however, lower rates may be attributable to increased activity during peak flu season, which may also suggest the need for higher staffing levels – further investigation is needed to uncover the reasons behind these associations.

*3.2.2 RReliefF*

In this subsection we discuss the results of RReliefF feature ranking obtained for each of the 10 facilities being investigated; the results are presented in Table 8.

| **Fac #** | *temp* | *humid* | *weekday* | *flu* | *holiday* | *night* | *July* |
|---|---|---|---|---|---|---|---|
| 147 | 2 | 5.7 | 3.8 | 1 | 5.2 | 4 | 6.3 |
| 155 | 1 | 2 | 4.8 | 4.8 | 5.2 | 3.2 | 7 |
| 153 | 2.2 | 5.8 | 4.5 | 1 | 4.7 | 3.1 | 6.7 |
| 149 | 4.3 | 5.1 | 1.2 | 2.1 | 6 | 6.6 | 2.7 |
| 123 | 1 | 4.1 | 5.4 | 3 | 4.3 | 7 | 3.2 |
| 127 | 2.8 | 3.5 | 6.5 | 4.4 | 1 | 3.3 | 6.5 |
| 91 | 3.9 | 3 | 5.5 | 2.1 | 1 | 5.5 | 7 |
| 101 | 3.1 | 7 | 5.3 | 4 | 3 | 4.6 | 1 |
| 170 | 1.9 | 4.4 | 3.9 | 1.2 | 4.3 | 6.9 | 5.4 |
| 168 | 1.5 | 3.8 | 2.9 | 1.5 | 6.3 | 6.6 | 1.8 |

Table 8: Facility-specific RReliefF feature rankings.

The first observation we wish to make is that there is no single feature that completely dominates the feature rankings among the different facilities. This suggests that facilities' compliance rates are affected differently by our selected features. However, we can also that some features are often ranked as being more important, while others as less important. For instance, *temp* is frequently one of the top three features, while *July* more often appears toward the bottom of the ranking. It is important to note here, however, that while *July*, *weekday*, *night*, and *holiday* appear toward the end of the feature ranking for some facilities, they appear towards the top for others. The *flu* feature also frequently appears in the top three feature rankings among facilities, while *humid* often appears somewhere near the middle of the rankings.

*3.2.3 Marginal Effects*

The facility-specific marginal effects modeling results are presented in Figure 5. Note that we are reporting only a subset of results, which include *temp*, *humid*, *weekday*, and *flu*.

Cumulatively, these results further support what we have already discussed, with a few observational caveats. First, temperature is found to be universally indicative of higher rates of compliance, which was found to not be entirely true for facilities 91 and 101; these coefficients are likely obscured by some degree of multicollinearity with other features – the same is true of *humid*. *weekday* and *flu*, as in the other results, are found to be mostly indicative of higher rates of compliance, with the exception of a few facilities.

| Facility | *temp* | *temp* | *weekday* | *flu* |
|---|---|---|---|---|
| 147 | | | | |
| 155 | | | | |
| 153 | | | | |
| 149 | | | | |
| 123 | | | | |
| 127 | | | | |
| 91 | | | | |
| 101 | | | | |
| 170 | | | | |
| 168 | | | | |

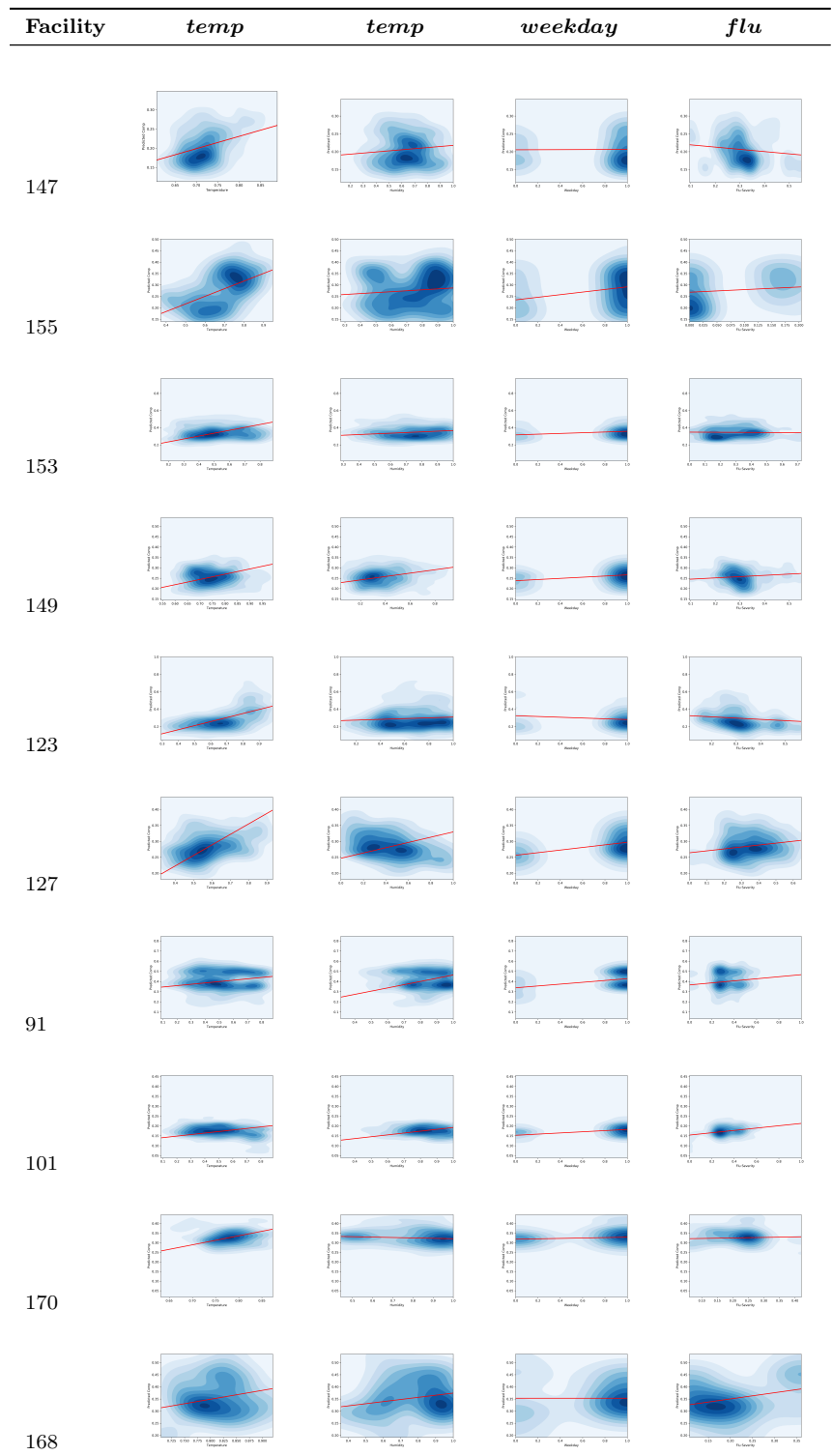

Fig. 5: Facility-specific marginal effects modeling results.

## 4 Discussion and Future Work

In this section we discuss the broader implications of our findings, as well as directions for future work.

The global results, including the full $M5$ Ridge Regression model, marginal effects models, and RReliefF feature ranking, provide several insights. First, we found that facility identities are strongly related to compliance, suggesting that facility-wide attitudes towards hand hygiene exist, persist in time, and are predictive of compliance rates. On the other hand, this observation may also reflect differences in sensor installation, where different facilities may have sensors instrumented in different departments, thus affecting reported rates. Second, increases in influenza severity were associated with an increase in compliance, which is encouraging because it implies that healthcare workers are responding positively (i.e, more hand hygiene) to an increased presence of infectious disease. This artifact also surfaced in our facility-specific models, which also revealed that different facilities have different magnitudes in the effect of flu severity on hand hygiene. Third, our conjecture regarding lower weekend and holiday compliance appears to have some merit, although the specific holidays associated with negative compliance were somewhat surprising. We again acknowledge that this result may be affected by increased visitors during these times, diluting the perceived compliance rate. Furthermore, our facility-specific modeling showed that, for some hospitals, both *weekday* and *holiday* had a large bearing on hand hygiene compliance predictions (i.e., these factors were important predictors of compliance). Fourth, our conjectures that higher humidity and temperature are indicative of higher rates of compliance were confirmed by the full model, marginal effects model, and statistical analysis. This finding is important as health care workers often cite skin irritation or dry skin as reasons for reduced frequency of hand hygiene. These same factors were also strongly suggested by our facility-specific modeling. Fifth, we found that compliance during the first week of residents' attendance ran contrary to our original conjecture: the *July* was essentially unobservable. However we did find that select facilities (153, 101, and 168) had this as an influencing factor (particularly 101 and 168). Finally, we found that *night* was associated with slightly lower compliance rates. However, as our facility-specific modeling exposed, some facilities (149, 123) appear to have slightly higher rates of compliance during the evening; although, it is worth noting that, for these facilities, *night* was at the bottom of the RReliefF feature ranking (indicating relatively low importance).

Different facilities have different factors that affect compliance rates differently: no two facilities are alike. While many of the facilities have factors that influence compliance rates in similar ways – positive or negative (e.g., temperature) – they differ in degree (how much these common factors influence compliance) and composition (the specific set of non-zero terms in the hypothesis vector $\mathbf{h}^*$). Cumulatively, we can see that factors affecting hand hygiene compliance among facilities is a complicated topic requiring further investigation.

This work has several limitations. First, there are differences among installations: not all doors and dispensers may be instrumented and, therefore, we cannot track, for example, the use of personal alcohol dispensers (we can only assume stable practices within facilities). Thus our compliance estimates may be based on partial information and are certainly not comparable across facilities. Second, our compliance estimates are facility wide, meaning that we do not exploit the co-location of dispensers and door event sensors, but only the temporal correlation of the individual events. Thus, our assumption that each door event corresponds to a hand-hygiene opportunity may be fundamentally flawed, even as it allows for consistent intra-facility comparisons. Third, we acknowledge the possibility of location and sampling bias with regard to both the sensors and facilities. If sensors were to be placed in only the ICU of one facility and in the emergency room of another, we may observe different rates, which may be entirely reasonable and expected in clinical practice. Additionally, though facilities are distributed across the United States, they are by no means meant to be a representative sample of facility types or climatic conditions.

In our future endeavors we would first like to consider alternative definitions of compliance and examine compliance at finer-grained temporal levels, perhaps exploring time-series analyses. We intend to also explore framing the problem as one of classification, rather than only regression, which may help tease out additional artifacts. Finally, data pertaining to compliance rates under certain interventions would give way to exploration of intervention efficacy both in general and using prediction-based methodology, such as inverse classification, to recommend facility-specific intervention policies [29, 30].

Hand hygiene compliance is a simple yet effective method of preventing the transmission of disease, both among the population at large, and within health care facilities, yet there have been few attempts to study the factors that can affect compliance. This study presents a first look at factors that underlie health care worker hand-hygiene compliance rates, including weather conditions, holidays and weekends, and infectious disease prevalence and severity, and serves as a model for future studies that will exploit the availability of temporally and spatially rich compliance data collected by the sophisticated sensor systems now being put into practice.

## 5 Conflicts of Interest

Philip M. Polgreen has received research funding from Company GOJO Industries, Inc. Author Jason Slater is an employee of GOJO Industries, Inc.

## References

1. R. Klevens, J. Edwards, C. Richards, and T. Horan, "Estimating health care-associated infections and deaths in us hospitals," *Public Health*, no. 122, pp. 160–166, 2007.
2. R. Roberts, R. Scott, B. Hota, L. Kampe, F. Abbasi, S. Schabowski, I. Ahmad, G. Ciavarella, R. Cordell, S. Solomon, R. Hagtvedt, and R. Weinstein, "Costs attributable to healthcare-acquired infection in hospitalized adults and a comparison of economic methods," *Medical Care*, vol. 48, no. 11, pp. 1026–1035, November 2010.
3. R. Roberts, B. Hota, I. Ahmad, R. Scott, S. Foster, F. Abbasi, S. Schabowski, L. Kampe, G. Ciavarella, M. Supino, J. Naples, R. Cordell, S. Levy, and R. Weinstein, "Hospital and societal costs of antimicrobial-resistant infection in a chiago teaching hospital: implications for antibiotic stewardship," *Clinical Infectious Diseases*, vol. 49, no. 8, pp. 1175–1184, October 2009.
4. J. M. Boyce and D. Pittet, "Guidelines for hand hygiene in health-care settings: recommendations of the healthcare infection control practices advisory committee and the hicpac/shea/apic/idsa hand hygiene task force," *Infection Control and Hospital Epidemiology*, no. 23, pp. S3–S41, 2002.

5. B. Allegranzi, H. Sax, L. Bengaly, H. Richet, D. Minta, M. Chraiti, F. Sokona, A. Gayet-Ageron, P. Bonnabry, and D. Pittet, "World health organization "point g" project management committee. successful implementation of the world health organization hand hygiene improvement strategy in a referral hospital in mali, africa," *Infection Control and Hospital Epidemiology*, vol. 31, no. 2, pp. 133–141, February 2010.
6. D. Pittet, B. Allegranzi, and J. Boyce, "World health organization world alliance for patient safety first global patient safety challenge core group of experts. the world health organization guidelines on hand hygiene in health care and their consensus recommendations," *Infection Control and Hospital Epidemiology*, vol. 30, no. 7, pp. 611–622, July 2009.
7. J. P. Hass and L. E. L., "Measurement of compliance with hand hygiene," *Journal of Hospital Infection*, no. 66, pp. 6–14, 2007.
8. J. Boyce and M. Cooper, T anda Dolan, "Evaluation of an electronic device for real-time measurement of alcohol-based hand rub use," *Infection Control and Hospital Epidemiology*, vol. 30, no. 11, pp. 1090–1095, November 2009.
9. Joint Commission of Accreditation of Healthcare Organizations, "Patient safety goals," Joint Commission of Accreditation of Healthcare Organizations, Tech. Rep., 2017. [Online]. Available: http://www.jcaho.org/accredited+organizations/patient+safety/npsg.htm
10. J. Fries, A. Segre, G. Thomas, T. Herman, K. Ellingson, and P. Polgreen, "Monitoring hand hygiene via human observers: How should we be sampling?" *Infection Control and Hospital Epidemiology*, vol. 33, no. 7, pp. 689–695, Jul. 2012, [PMID: 22669230].
11. D. Sharma, G. Thomas, E. Foster, J. Iacovelli, K. Lea, J. Streit, and P. Polgreen, "The precision of human-generated hand-hygiene observations: a comparison of human observation with an automated monitoring system," *Infection Control and Hospital Epidemiology*, vol. 33, no. 12, pp. 1259–1261, December 2012.
12. T. Eckmanns, J. Bessert, M. Behnke, and H. Gastmeier, P anda Ruden, "Compliance with antiseptic hand rub use in intensive care units: The hawthorne effect," *Infection Control and Hospital Epidemiology*, no. 27, pp. 931–934, 2006.
13. M. Monsalve, S. Pemmaraju, G. Thomas, T. Herman, and P. Segre, AM anda Polgreen, "Do peer effects improve hand hygiene adherence among healthcare workers?" *Infection Control and Hospital Epidemiology*, vol. 35, no. 10, pp. 1277–1285, October 2014.
14. V. Boscart, K. McGilton, A. Levchenko, G. Hufton, P. Holliday, and G. Fernie, "Acceptability of a wearable hand hygiene device with monitoring capabilities," *Journal of Hospital Infection*, vol. 70, no. 3, pp. 216–222, November 2008.
15. A. Venkatesh, M. Lankford, D. Rooney, T. Blachford, C. Watts, and G. Noskin, "Use of electronic alerts to enhance hand hygiene compliance and decrease transmission of vancomycin-resistant enterococcus in a hematology unit," *American Journal of Infection Control*, vol. 36, no. 3, pp. 199–205, April 2008.
16. P. M. Polgreen, C. S. Hlady, M. a. Severson, A. M. Segre, and T. Herman, "Method for automated monitoring of hand hygiene adherence without radio-frequency identification." *Infection control and hospital epidemiology : the official journal of the Society of Hospital Epidemiologists of America*, vol. 31, no. 12, pp. 1294–1297, 2010.
17. M. T. Lash, J. Slater, P. M. Polgreen, and A. M. Segre, "A large-scale exploration of factors affecting hand hygiene compliance using linear predictive models," in *Healthcare Informatics, 2017 IEEE International Conference on (ICHI)*, 2017, pp. 66–73. [Online]. Available: http://ieeexplore.ieee.org/document/8031133/
18. H. Dai, K. L. Milkman, D. A. Hofmann, and B. R. Staats, "The Impact of Time at Work and Time Off from Work on Rule Compliance: The Case of Hand Hygiene in Healthcare," *Journal of Applied Psychology*, vol. 100, no. 3, pp. 846–862, 2014. [Online]. Available: http://papers.ssrn.com/sol3/papers.cfm?abstract_id=2423009
19. C. Jarrin Tejada and G. Bearman, "Hand Hygiene Compliance Monitoring: the State of the Art," *Current Infectious Disease Reports*, vol. 17, no. 4, 2015. [Online]. Available: http://link.springer.com/10.1007/s11908-015-0470-0
20. E. Kalnay, M. Kanamitsu, R. Kistler, W. Collins, D. Deaven, L. Gandin, S. Iredell, S. Saha, G. White, Y. Zhu, a. Leetmaa, R. Reynolds, M. Chelliah, W. Ebisuzaki, W. Higgins, J. Janowiak, K. Mo, C. Ropelewski, J. Wang, R. Jenne, and D. Joseph, "The NCEP/NCAR 40-Year Reanalysis Project," pp. 437–471, 1996. [Online]. Available:
21. N. R. Draper, H. Smith, and E. Pownell, *Applied Regression Analysis*.  Wiley New York, 1966, vol. 3.
22. J. R. Quinlan, "Learning with continuous classes," in *5th Australian Joint Conference on Artificial Intelligence*, vol. 92, 1992, pp. 343–348.

23. F. D. Johansson, U. Shalit, and D. Sontag, "Learning representations for counterfactual inference," in *33rd International Conference on Machine Learning (ICML)*, 2016.
24. R. Tibshirani, "Regression shrinkage and selection via the lasso," *Journal of the Royal Statistical Society. Series B (Methodological)*, pp. 267–288, 1996.
25. H. Zou and T. Hastie, "Regularization and variable selection via the elastic net," *Journal of the Royal Statistical Society: Series B (Statistical Methodology)*, vol. 67, no. 2, pp. 301–320, 2005.
26. M. Robnik-Šikonja and I. Kononenko, "An adaptation of relief for attribute estimation in regression," in *Machine Learning: Proceedings of the Fourteenth International Conference (ICML97)*, 1997, pp. 296–304.
27. K. Kira and L. A. Rendell, "A practical approach to feature selection," in *Proceedings of the ninth international workshop on Machine learning*, 1992, pp. 249–256.
28. R. Williams *et al.*, "Using the margins command to estimate and interpret adjusted predictions and marginal effects," *The Stata Journal*, vol. 12, no. 2, p. 308, 2012.
29. M. T. Lash, Q. Lin, W. N. Street, J. G. Robinson, and J. Ohlmann, "Generalized inverse classification," in *Proceedings of the 2017 SIAM International Conference on Data Mining (SDM'17)*, 2017, pp. 162–170. [Online]. Available: https://doi.org/10.1137/1.9781611974973.19
30. M. T. Lash, Q. Lin, W. N. Street, and J. Robinson, "A budget constrained inverse classification framework for smooth classifiers," in *Data Mining Workshops (ICDMW), 2017 IEEE International Conference on*, 2017, pp. 1184–1193.